\documentclass[cits]{PoS}

\title{An RMHD study of transition between prompt and afterglow GRB phases}

\ShortTitle{An RMHD study of transition between prompt and afterglow GRB phases}

\author{\speaker{Petar Mimica}\\
        Department for Astronomy and Astrophysics, University of Valencia\\
        E-mail: \email{petar.mimica@uv.es}}

\author{Dimitrios Giannios\\
        Max-Planck Institute for Astrophysics\\
        E-mail: \email{giannios@mpa-garching.mpg.de}}

\author{Miguel-Angel Aloy\\
        Department for Astronomy and Astrophysics, University of Valencia\\
        E-mail: \email{miguel.a.aloy@uv.es}}

      \abstract{We study the afterglow phases of a GRB through
        relativistic magnetohydrodynamic simulations. The evolution of
        a relativistic shell propagating into a homogeneous external
        medium is followed.  We focus on the effect of the
        magnetization of the ejecta on the initial phases of the
        ejecta-external medium interaction. In particular we are
        studying the condition for the existence of a reverse shock
        into the ejecta, the timescale for the transfer of the energy
        from the shell to the shocked medium and the resulting
        multiwavelength light curves. To this end, we have developed a
        novel scheme to include non-thermal processeses which is
        coupled to the relativistic magnetohydrodynamic code
        \emph{MRGENESIS} in order to compute the non-thermal
        synchrotron radiation.}

\FullConference{Supernovae: lights in the darkness\\
		 October 3-5, 2007\\
		 Maó (Menorca) }

\begin{document}

\section{Introduction}

Magnetic fields may play an important role in the relativistic flow
of a gamma-ray burst (GRB), but the extent to which they are
important remains uncertain. Looking at the process which produces a
relativistic GRB outflow, two alternatives are usually considered. On
the one hand, neutrino annihilation may be a process which leads to
the formation of fireball dominated by thermal energy. Here magnetic
fields are dynamically unimportant. Alternatively, powerful enough
magnetic fields can efficiently extract the rotational energy from the
central engine and launch a Poynting-flux dominated flow.

In order to achieve relativistic velocities, GRBs have to be launched
with high energy-to-mass ratio. In the fireball model, most of the
energy is assumed to be thermal \cite{G86,P86}. This can be
be result of neutrino-antineutrino annihilation in the polar region of
an accreting object \cite{W93,Aetal00,AJM05,Betal07}. The acceleration of the
fireball is due to the internal pressure gradient, whereby thermal
energy of the fireball is converted into kinetic energy of the
baryons. At the end of the acceleration phase faster parts of the flow
may collide with the slower ones producing the internal shocks which
power the GRB prompt emission \cite{RM94,DM98,MAMB05,MAM07}. Processes
such as two-stream instability in the shocks might amplify weak
magnetic fields present in the flow \cite{ML99}. These fields are
expected to account for less than $1$\% of the energy of the
flow. After the internal shock phase the flow expands and cools as it
enters the afterglow phase. In this case very weakly magnetized flow
is expected at the onset of the afterglow.

Magnetic fields with appropriate topology can efficiently extract
rotational energy from the central engine, be it an accretion disk
\cite{BP82}, rotating black hole \cite{BZ77} or a millisecond magnetar
\cite{U92} launching a Poynting-flux dominated flow (PDF). The
acceleration of the PDF depends on the field geometry and the
dissipation processes. Magnetic dissipation can convert Poynting flux
into kinetic energy \cite{DS02,GS06} and also power the GRB prompt
emission \cite{LB03,G06,GS07}. Different studies of MHD jet
acceleration show that magnetic energy is not completely converted
into kinetic energy at the end of the acceleration phase. At larger
radii we expect magnetic energy of the flow to be comparable to the
kinetic energy of the baryons \cite{GS06} or even much larger
\cite{LB03,T06}. It should be noted here that the magnetization of the
flow might be decreased by the pair-loading caused by the
$\nu\bar{\nu}$-annihilation near the central engine \cite{LE93, OA07}.

Fireball and PDF models respectively predict weakly and strongly
magnetized flow at the onset of the afterglow phase. The initial
phases of the interaction of the GRB flow with the (circumburst)
external medium depend on the strength of the magnetic fields in the
flow. A particularly promising probe of the magnetization of the GRB
flow can come from understanding the early afterglow emission
\cite{KP03,ZKM03,GMA07}.

In this paper we outline the status of the ongoing study of the
interaction of magnetized ejecta with external medium. In
Sec.~\ref{sec:interaction} we describe in more detail the current
understanding of ejecta-medium interaction. Sec.~\ref{sec:simulations}
gives an overview of numerical methods and a plan for numerical
simulations. Summary is given in Sec.~\ref{sec:conclusions}.

\section{Ejecta-medium interaction}
\label{sec:interaction}

We consider a homogeneous shell expanding into an external medium of
constant density\footnote{Similar analysis can be performed for the
  wind profile where the density of the external medium scales as
  $r^{-2}$.}. At large distances from the central engine (typically
$R_0\approx 10^{15} - 10^{17}$ cm) substantial interaction begins,
whereby ejecta begins to decelerate due to the accumulation of the
external material. We assume that at these distances the flow has
already been accelerated and collimated. The internal dissipation
mechanism, presumably responsible for the prompt emission
(e.g. internal shocks, magnetic dissipation) is also expected to take
place at a shorter distance from the central engine with respect to
that of the afterglow phases. After the internal dissipation is over,
the flow expands and cools. Since we are interested in the afterglow
phases, the shell is assumed to be cold. We denote the shell Lorentz
factor by $\gamma_0\gg 1$ and its width by $\Delta_0$. In a radially
expanding outflow the magnetic field component perpendicular to the
direction of motion drops as $r^{-1}$ while the component in the
direction of motion drops as $r^{-2}$, so that we expect the magnetic
field to be dominated by the perpendicular component. We define the
magnetization parameter as
\begin{equation}
\label{eq:magnetization}
\sigma_0 := \frac{E_P}{E_K} = \frac{B_0^2}{4\pi \gamma_0 \rho_0 c^2}\, ,
\end{equation}
where $E_P$ and $E_K$ are Poynting and kinetic energies in the shell,
$\rho_0$ and $B_0$ its density and magnetic field measured in the
central engine frame. With this definition a fireball corresponds to
$\sigma_0\ll 1$ while a PDF has $\sigma_0\ge 1$. $c$ is the speed of
light. The total energy of the shell is
\begin{equation}
\label{eq:totenergy}
E = 4\pi R_0^2\Delta_0(\gamma_0\rho_0 c^2 + B_0^2/4\pi) = E_K(1+\sigma_0)\, .
\end{equation}
From Eqs.~\ref{eq:magnetization} and \ref{eq:totenergy} we can see
that $\sigma_0$ parametrizes the fraction of the total energy in the
kinetic ($1/(1+\sigma_0)$) and magnetic ($\sigma_0/(1+\sigma_0)$)
form. 

We first discuss the ejecta-medium interaction for non-magnetized
ejecta, and then turn to the arbitrarily magnetized case. We also
focus on the conditions for the existence of a reverse shock into
ejecta of arbitrary magnetization.

\subsection{The case of $\sigma_0\ll 1$}

The evolution of the interface between the cold unmagnetized shell and
the external medium is well understood. It was studied in detail
analytically \cite{SP95,MR97} as well as using one-dimensional
\cite{KPS99} and two-dimensional \cite{CGV04,Getal01,Metal07}
numerical simulations.

At the interface between the shell and the ambient medium two shocks
form, the \emph{forward} shock propagating into the external medium,
and the \emph{reverse} shock propagating into the shell. Shocked shell
and external medium are separated by the contact discontinuity. The
forward shock is always ultra-relativistic, while the strength of the
reverse shock depends on the density contrast between the shell and
the external medium and the bulk Lorentz factor $\gamma_0$. We
distinguish between \emph{relativistic} and \emph{Newtonian} reverse
shocks \cite{SP95}. The critical parameter is
\begin{equation}
\label{eq:ksi}
\xi := l^{1/2}\Delta_0^{-1/2}\gamma_0^{-4/3}\, ,
\end{equation}
where $l=(3E/4\pi n_e m_p c^2)^{1/3}$ is the Sedov length, $n_e$ the
external medium number density, and $m_p$ the proton mass. In the
Newtonian case ($\xi\gg 1$) the shock is non-relativistic in the shell
rest frame and does not decelerate the ejecta much, rather the ejecta
decelerate once they accumulate a mass $\gamma_0^{-1}$ times their own
mass from the external medium. In the relativistic case ($\xi\ll 1$)
the shock crosses the ejecta quickly and slows them down
considerably. After the shock crosses the ejecta, there is a phase
where shocks and rarefaction waves cross the ejecta, passing the
energy to the forward shock. At later stages the evolution of the
ejecta only depends on their total energy and the external medium
density \cite{BM76}.

\subsection{The case of arbitrary $\sigma_0$}

The dynamics of magnetized ejecta has not been studied as thoroughly
as that of unmagnetized ejecta. A qualitative difference to the
unmagnetized case is that later phases of the evolution are influenced
by the internal evolution of the magnetized shell. The initial phase
of the evolution has recently been studied \cite{ZK05} by solving the
ideal MHD shock conditions for arbitrarily magnetized ejecta with
toroidal field. In particular, the dynamics of shock crossing has been
studied assuming that there is a reverse shock. In that case, the
reverse shock crosses the shell faster the higher the magnetization
is. However, as we have recently shown \cite{GMA07}, it is not always
the case that a reverse shock forms.

\subsection{Conditions for the existence of a reverse shock}
\label{sec:cond}

Cold, non-magnetized ejecta are always crossed by a reverse shock upon
interacting with the external medium. This is the case since the sound
speed of the ejecta is low and does not allow for fast transfer of the
information of the interaction with the external medium throughout
their volume.  On the other hand, in a flow that is strongly
magnetized and sub-fast magnetosonic (as in the Lyutikov \& Blandford
2003 model \cite{LB03}) there is no reverse shock forming. The flow
adjusts gradually to the changes of the pressure in the contact
discontinuity that separates the magnetized flow from the shocked
external medium.  In \cite{GMA07} we generalize to arbitrarily
magnetized ejecta and derive the condition for the formation of a
reverse shock.

After a detailed treatment of this problem we arrive to the following
condition for the formation of a reverse shock \cite{GMA07}
\begin{equation}
  \label{eq:xisigma}
  \xi < \frac{1}{(4\sigma_0)^{1/3}}\, ,
\end{equation}
which can be rewritten in terms of shell parameters as\footnote{We use
  the convention that $A = A_x 10^x$.}
\begin{equation}
  \label{eq:sigmaprm}
  \sigma_0 <0.6 n_0^{1/2}\Delta_{12}^{3/2}\gamma_{2.5}^4E_{53}^{-1/2}\, .
\end{equation}

\begin{figure}
  \begin{center}
    \includegraphics[scale=0.6, angle=270]{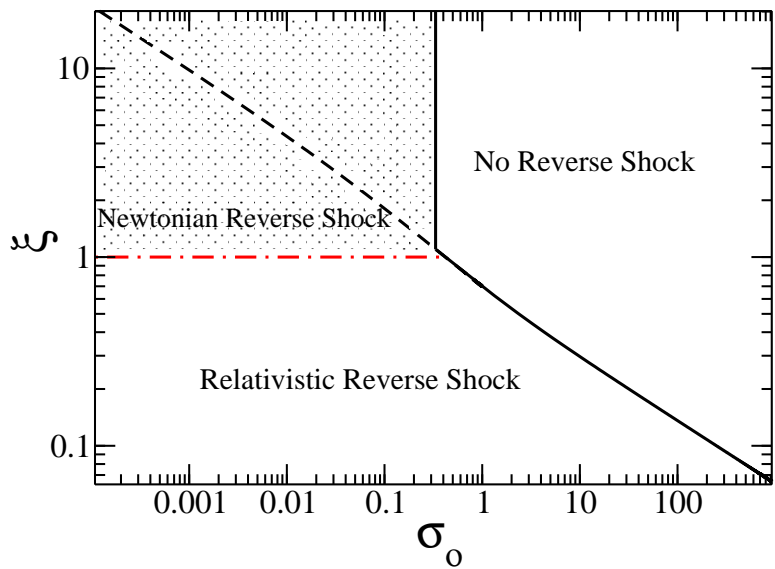}
    \caption{Existence of a reverse shock in the $\xi-\sigma_0$
      parameter space. The dashed black line delimits regions where a
      reverse shock forms from the region where there is no reverse
      shock, ignoring the radial shell spreading. The solid black line
      shows the delimitation when the shell spreading is taken into
      account. See \cite{GMA07} for details.}
  \end{center}
  \label{fig:xisigma}
\end{figure}

Fig.~1 (taken from \cite{GMA07}) shows the division of $\xi-\sigma_0$
parameter space in two regions, one where a reverse shock forms and
another where its formation is suppressed. We note that from the
conditions in Eqs.~\ref{eq:xisigma} and \ref{eq:sigmaprm} it follows
that even for mildly magnetized shells a reverse shock can be
suppressed. This indicates that the paucity of the observed optical
flashes in GRB afterglows (associated with the reverse shock emission)
may be caused by the suppression of the shock in many GRBs.

\section{Numerical simulations}
\label{sec:simulations}

In this section we describe the reasons and motivation for performing
numerical simulations of shell-ejecta interactions:

\begin{enumerate}

\item {\bf Verification of the analytic approach:} Results of the
  analytic work described in Sec.~\ref{sec:interaction}, especially in
  Sec.~\ref{sec:cond}, need to be verified by means of numerical
  simulations. We note that the line dividing regions of formation and
  suppression of the reverse shock in Fig.~1, given by
  Eq.~\ref{eq:xisigma} is approximate, and it is necessary to perform
  numerical simulations for models whose initial parameters lie in the
  vicinity of the line.

\item {\bf Dynamics of shock propagation:} We want to use numerical
  models to study the influence of the magnetization on the
  propagation of the reverse shock through the shell. On Figs.~2 and
  3 we show snapshots of two simulations, one with the unmagnetized
  ($\sigma_0=0$), and another with the magnetized ($\sigma_0=1$) shell
  interacting with the external medium, both shown at the same
  evolutionary time. It can be seen that in the magnetized case
  (Fig.~3) the reverse shock has penetrated the shell deeper than in
  the unmagnetized case (Fig.~2). Our goal is to perform a parametric
  study where we study the propagation of shocks and rarefactions
  through the shell for different combinations of $\xi$ and
  $\sigma_0$.

\item {\bf Long term evolution and energy content:} One of the
  important questions that simulations can answer is the timescale of
  the transfer of energy from the shell to the shocked external
  medium. Long-term numerical calculations are needed to determined
  the dependence of the efficiency of the energy transfer on
  magnetization. We also want to investigate the long-term evolution of the
  blast wave and determine when the shell profile relaxes to the
  Blandford-McKee solution \cite{BM76}.

\item {\bf Light curves:} Finally, we wish to compute synthetic
  multi-wavelength afterglow light curves. Light curves are very
  sensitive to the magnetic field content of the shell, shock strength
  and, especially, detailed radial profile $\gamma(R)$ of the Lorentz
  factor as the shell propagates into the external medium. They also
  depend on the distribution of shock accelerated particles. To see why
  $\gamma(R)$ is crucial for the light curve, consider the difference
  in the arrival time to the observer of signals emitted
  simultaneously from two points with radii $R$ and $R+\mathrm{d}R$,
  respectively. It turns out that the difference is $\mathrm{d}t
  \approx \mathrm{d}R\ \gamma(R)^{-2}$ for a relativistic shell. As
  can be seen, a sudden drop in a Lorentz factor in one model as
  compared to the other will produce features which have longer
  observed duration. The resulting light curves of two models can be
  very different. High-resolution simulations are needed to resolve
  sufficiently short time intervals in order to be able to study the
  influence of magnetization on $\gamma(R)$ and the subsequent light
  curve.

\end{enumerate}

\begin{figure}
  \begin{center}
    \includegraphics[scale=0.4]{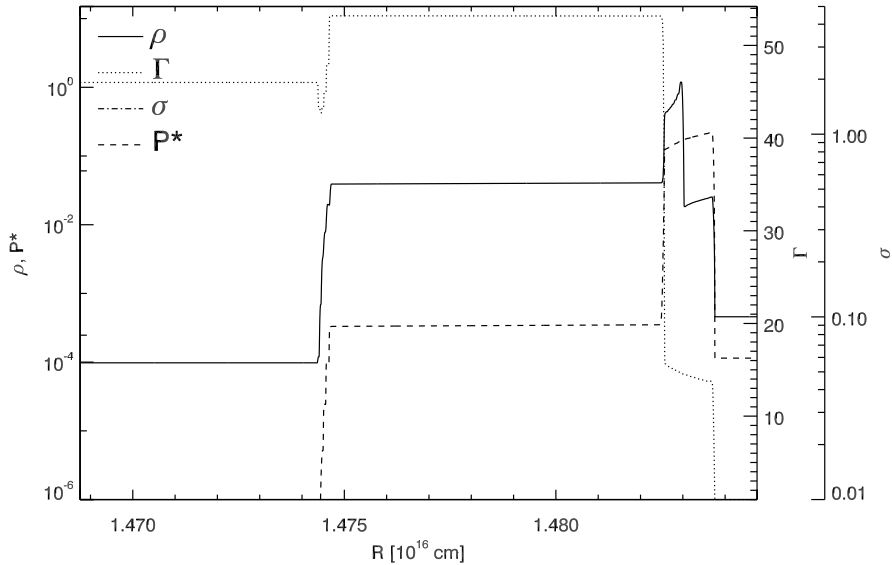}
    \caption{Snapshot from relativistic hydrodynamical simulation of a
      spherical, non-magnetized ($\sigma=0$) shell that decelerates
      interacting with external medium of density $10$ cm$^{-3}$. The
      total energy of the shell is $E=10^{51}$ erg, its initial width
      $\Delta_0=10^{14}$ cm and bulk Lorentz $\gamma_0\simeq
      50$. $P^*$ (dashed line), $\rho$ (solid line) stand for the gas
      pressure and (lab frame) density respectively (in arbitrary
      units). With increasing radius, one can clearly see the reverse
      shock, contact discontinuity and forward shock located at $r\sim
      1.483\cdot 10^{16}$ cm.}
  \end{center}
\end{figure}

\begin{figure}
  \begin{center}
    \includegraphics[scale=0.4]{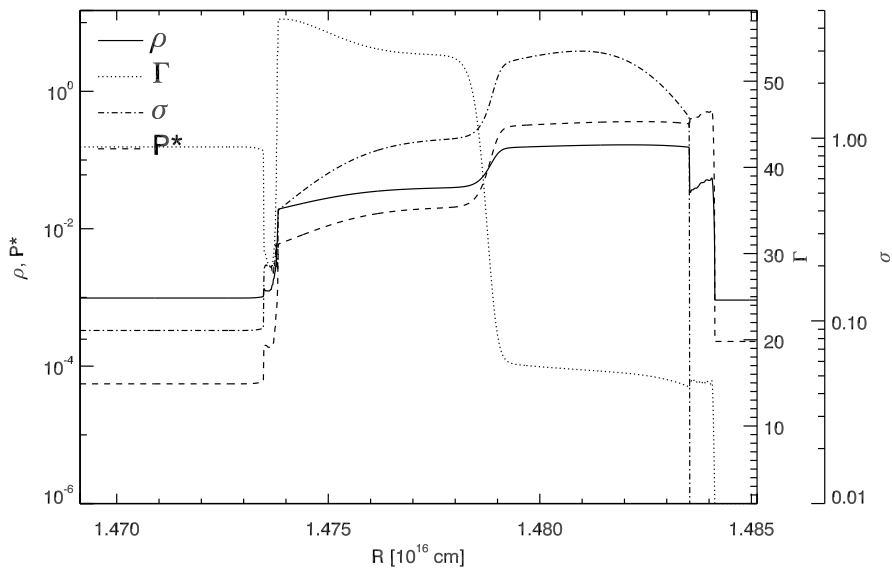}
    \caption{Same as Fig.~2, but for the magnetized ($\sigma_0=1$)
      shell. Note that the reverse shock crosses the ejecta faster
      with respect to the non-magnetized case in agreement with
      analytical expectations.}
  \end{center}
\end{figure}

We have developed a relativistic magnetohydrodynamic code
\emph{MRGENESIS} \cite{MAMB04,MAMB05,MAM07} which consists of a
finite-volume, high-resolution, shock-capturing scheme \emph{GENESIS}
\cite{A99,API99,L05} which solves for the conservation laws of
relativistic magnetohydrodynamics, and a module which follows the
transport, evolution and radiation from non-thermal particles. For the
purpose of analyzing dynamics and emission from afterglow shells, a
very high resolution is needed. Numerical convergence tests have shown
that, in order to sufficiently resolve the shell, we need to resolve
the scales of the order of $\Delta_0 \gamma_0^{-2}$. This means that
we need to use at least $\gamma_0^2$ zones within the shell. We use a
grid re-mapping procedure described in \cite{MAMB04}, which can be
thought of as a guided mesh refinement centered on the shell. Even
with this procedure the actual number of zones for a typical model is
expected to be several millions. We are performing simulations on
supercomputers of the Spanish Supercomputing Network.

\section{Conclusions}
\label{sec:conclusions}

We are performing high-resolution numerical studies of the transition
from prompt to the early afterglow phase of gamma-ray bursts. The
afterglow is being modeled as the radiation from the relativistic
shell expanding into the homogeneous external medium. We study the
difference between the fireball (unmagnetized shell) and the
Poynting-flux dominated (magnetized shell) models. 

In the context of the early optical afterglow we have shown
analytically that even a moderate magnetization of the flow can
suppress the existence of a reverse shock, and thus explain the
apparent paucity of the optical flashes for a large number of early
afterglows. We are currently performing simulations to study the
formation and suppression of relativistic shocks in detail. To study
the later phases of the afterglow we aim to determine the influence of
the initial shell magnetization on the energy content, transfer of
energy from the shell to the forward shock, and the long-term flow
structure.

We have developed a novel scheme for treating non-thermal processes in
relativistic magnetohydrodynamic simulations. This scheme is used
to compute multi-wavelength light curves from numerical
simulations. The aim is to study the influence of the initial
magnetization on the short- and long-term light curves.

\section*{Acknowledgements}
PM is at the University of Valencia with a European Union Marie Curie
Incoming International Fellowship (MEIF-CT-2005-021603).  MAA is a
Ram\'on y Cajal Fellow of the Spanish Ministry of Education and
Science. He also acknowledges partial support from the Spanish
Ministry of Education and Science (AYA2004-08067-C03-C01,
AYA2007-67626-C03-01). The authors thankfully acknowledge the
computer resources, technical expertise and assistance provided by the
Barcelona Supercomputing Center - Centro Nacional de Supercomputación.

\end{document}